\documentclass[aps,prl,preprint,amsmath,amssymb,showpacs]{revtex4}

\usepackage{graphicx}

\def\bc{\begin{center}}
\def\ec{\end{center}}
\def\be{\begin{eqnarray}}
\def\ee{\end{eqnarray}}

\begin{document}


\title{Unification of Dark Matter and Dark Energy \\in a Modified Entropic Force Model}



\author{Zhe Chang$^{1,2}$\footnote{E-mail: changz@ihep.ac.cn}}
\author{Ming-Hua Li$^{1,2}$\footnote{E-mail: limh@ihep.ac.cn}}
\author{Xin Li$^{2,3}$\footnote{E-mail: lixin@itp.ac.cn}}
\affiliation{${}^1$Institute of High Energy Physics, Chinese Academy of Sciences, 100049 Beijing, China\\
${}^2$Theoretical Physics Center for Science Facilities, Chinese Academy of Sciences\\
${}^3$Institute of Theoretical Physics, Chinese Academy of Sciences, 100190 Beijing, China}



\begin{abstract}
  In Verlinde's entropic force scenario of gravity, Newton's laws and Einstein equations can be obtained from the first pinciples and general assumptions. However, the equipartition law of energy is invalid at very low temperatures. We show clearly that the threshold of the equipartition law of energy is related with horizon of the universe. Thus, a one-dimension Debye (ODD) model in the direction of radius of the modified entropic force (MEF) maybe suitable in description of the accelerated expanding universe. We present a Friedmann cosmic dynamical model in the ODD-MEF framework. We examine carefully constraints on the ODD-MEF model from the Union2 compilation of the Supernova Cosmology Project (SCP) collaboration, the data from the observation of the large-scale structure (LSS) and the cosmic microwave background (CMB), i.e. SNe Ia+LSS+CMB. The combined numerical analysis gives the best-fit value of the model parameters $\zeta\simeq10^{-9}$ and $\Omega_{m0}=0.224$, with $\chi_{min}^2=591.156$. The corresponding age of the universe agrees with the result of D.~Spergel {\it et al.}\cite{Spergel2003} at $95\%$ confidence level. The numerical result also yields an accelerated expanding universe without invoking any kind of dark energy. Taking $\zeta(\equiv 2\pi \omega_D/H_0)$ as a running parameter associated with the structure scale $r$, we obtain a possible unified scenario of the asymptotic flatness of the radial velocity dispersion of spiral galaxies, the accelerated expanding universe and the Pioneer 10/11 anomaly in the entropic force framework of Verlinde.
\end{abstract}

\pacs{95.36.+x,95.35.+d,98.80.-k}

\maketitle

\section{1. Introduction}
     The law of gravity, which was first discovered by Isaac Newton in the 18th century and later reformulated by Albert Einstein in the early 20th century, still remains unclear in the microscopic level nowadays. Enlightened by Hawking {\it et al.}'s work\cite{Hawking, Bekenstein, Hawking2} about the black hole entropy in 1970's, Jacobson\cite{Jacobson} got the gravitational field equations as Einstein's, starting from the first law of thermodynamics. Further research of Padmanabhan\cite{Padmanabhan1,Padmanabhan2} also gives gravity a thermodynamical interpretation. These results prompt people to take a statistical physics point of view on gravity. Recently, Verlinde\cite{Verlinde} reinterpreted gravity as an entropic force caused by a change of amount of information associated with the positions of bodies of matter. Newton's second law could be obtained with the introduction of Unruh temperature\cite{Unruh}. He also got the Newton's law of gravity and the Komar's definition of mass\cite{Komar} in a static curved space in relativistic case from the holographic principle\cite{Susskind} and black hole thermodynamics.

     Lots of work have been done to reveal the implications of the entropic force interpretation of gravity. To name a few: derivation of Friedmann equation in the entropic force framework\cite{Shu,Cai1,Cai3,Cai4}, the corresponding Newton gravity formulation in loop quantum gravity\cite{Smolin}, the construction of holographic actions from black hole entropy\cite{Makela,Caravelli}, the entropic force scenario of holographic dark energy\cite{LiM}, the generalization of the Newton's potential in the relativistic case\cite{Tian}, and the modified entropic force(MEF) due to the modification of the equipartition law of energy at very low temperature\cite{Gao}, etc.

     In Verlinde's paper\cite{Verlinde}, the equipartition law of energy of free particles plays a key role in the derivation of Newton's laws. Gao\cite{Gao} pointed out that the equipartition law of energy does not hold at very low temperatures. He made a modification simply using the three-dimension Debye model in solid state physics. His work provides an explanation for the accelerated expanding universe without invoking dark energy. However, we note that only the vibration in the direction of radius is related with an observable quantity and the ``low temperature'' should be fixed by a threshold. So that, a one-dimension Debye(ODD) function is suitable to be used to revise the free particle's equipartition law of energy in the study of cosmology\cite{Xin Li}. In fact, such a modification leads us to the famous modification of Newtonian dynamics(MOND). MOND was constructed to account for the asymptotic flatness of the rotational velocity curves of spiral galaxies\cite{Trimble,Rubin,TF}.

     In this paper, we show clearly that the threshold of equipartition law of energy is related with horizon of the universe. A Friedmann cosmic dynamical model is set up in the framework of the ODD-MEF. We examine carefully constraints on the ODD-MEF model from the Union2 compilation of the Supernova Cosmology Project(SCP) collaboration, the data from the observation of the large-scale structure(LSS) and the cosmic microwave background(CMB), i.e. SNe Ia+LSS+CMB. Results yield an accelerated expanding universe without invoking any kind of dark energy. Furthermore, by taking $\zeta(\equiv 2\pi \omega_D/H_0)$ as a running parameter associated with the structure scale $r$, we obtain a possible unified scenario of the asymptotic flatness of the radial velocity dispersion of spiral galaxies, the accelerated expanding universe and the Pioneer 10/11 anomaly in the entropic force framework of Verlinde.

     The paper is organized as follows. In section 2, a brief review on Verlinde's work and the connection between the ODD-MEF model and MOND is presented. We derive the corresponding Friedmann equation and consider the cosmological constraints on it in setion 3. In section 4, an alternative approach to the Pioneer 10/11 anomaly is presented. In section 5, by taking $\zeta$ as a running parameter associated with the structure scale $r$, we suggest a unified scenario of the asymptotic flatness of the rotation curves of spiral galaxies, the accelerated expanding universe and the Pioneer 10/11 anomaly in the framework of the ODD-MEF. Conclusions and further discussions can be found in section 6.

\section{2. From Modified Entropic Force To Mond}
     Think of a closed holographic screen and a free particle of mass $m$ near it on the side that spacetime has already emerged. The particle moves towards the screen, traveling a distance $\triangle x$ before merging into it. According to Bekenstein\cite{Bekenstein}, the change of entropy of the screen associated with the amount of information stored on it is
\begin{equation}
\label{entrophy}
\triangle S=2\pi k_B \frac{mc}{\hbar}\triangle x\ ,
\end{equation}
where $k_B$ is the Boltzmann constant and $c$ denotes the speed of light.
    The particle will experience an effective macroscopic force due to the statistical tendency to increase its entropy. This is described by
\begin{equation}
\label{entropic F}
F\triangle x=T\triangle S\ ,
\end{equation}
where $T$ is the temperature of the screen.
By introducing Unruh's relation between acceleration and temperature\cite{Unruh}
\begin{equation}
\label{Unruh T}
k_BT=\frac{1}{2\pi}\frac{\hbar a}{c}\ ,
\end{equation}
one recovers the second law of Newton
\begin{equation}
\label{Newton2}
F = ma\ .
\end{equation}

Suppose the holographic screen has a radius $R$, the area of which is $A = 4\pi R^2$. In theory of emergent space, each fundamental bit occupies one unit cell, the area of which on the screen is defined as $L_{p} \equiv \sqrt{G\hbar}$. $L_{p}$ is the Planck length. $G$ is mere a new constant here, which is later identified with Newton's constant. Let's denote the number of bits by $N$, which is given as
\begin{equation}
\label{bits}
N\equiv\frac{Ac^3}{L_p^2}= \frac{Ac^3}{G\hbar}= \frac{4\pi R^2 c^3}{G\hbar}\ .
\end{equation}
Each bit represents a microscopic degree of freedom. The total energy of the screen is given by the equipartition law
\begin{equation}
\label{equipartition}
E=\frac{1}{2}Nk_BT\ .
\end{equation}
Assuming that the energy of the screen is proportional to the mass $M$ that would emerge in the part of space enclosed by the screen itself as
\begin{equation}
\label{energy mass}
E=Mc^2\
\end{equation}
and using the equation (\ref{Newton2}), (\ref{bits}), (\ref{equipartition}) and (\ref{energy mass}), one obtains the familiar law\cite{Verlinde}:
\begin{equation}
\label{newton's gravity}
 F=G\frac{Mm}{R^2}\ .
\end{equation}

Thermodynamics and statistical physics tells us that the equipartition law of energy is valid only when the temperature is not very low. On the other hand, the Debye model in solid state physics is found to be successful in describing experimental results for most of the solid objects at very low temperatures\cite{Debye}. It should be noticed that the threshold of low energy for each direction of vibration is different, because it is related with the structure scale in the radial direction. In the case of astronomy and cosmology, the only temperature threshold is related with horizon of the universe. Thus, we focus on the ODD-MEF model throughout this paper. The equipartition law of energy can be rewritten as
\begin{equation}
\label{modified equipartition}
E=\frac{1}{2}Nk_BT\mathfrak{D}(x)\ ,
\end{equation}
where the one-dimension Debye function is defined as
\begin{equation}
\label{Debye function}
\mathfrak{D}(x)\equiv\frac{1}{x}\int^x_0\frac{y}{e^y-1}dy\ .
\end{equation}
$x$ is related to the Debye frequency $\omega_D$ and defined as
\begin{equation}
\label{x}
x\equiv\frac{\hbar\omega_D}{k_BT}=\frac{2\pi c\omega_D}{a}\ .
\end{equation}
Combining the equations (\ref{Newton2}), (\ref{bits}), (\ref{energy mass}) and (\ref{modified equipartition}), we obtain the modified Newton's law of gravitation
\begin{equation}
\label{modified gravity}
\frac{GM}{R^2}=a\mathfrak{D}\left(\frac{2\pi c\omega_D}{a}\right)\ .
\end{equation}

There are two limit cases for the equation (\ref{modified gravity}). One is the high temperature limit, with $x\ll1$. We have
\begin{equation}
\mathfrak{D}(x)\approx\frac{1}{x}\int^x_0dy=1.
\end{equation}
Therefore, the modified equipartition law of energy returns to the equation (\ref{equipartition}), with which Newton's law of gravity the equation (\ref{newton's gravity}) is recovered.

The other is the low temperature limit, with $x\gg1$. The Debye function $\mathfrak{D}(x)$ reduces to
\begin{equation}
\mathfrak{D}(x)\approx\frac{1}{x}\int^\infty_0\frac{y}{e^y-1}dy=\frac{\pi^2}{6x}.
\end{equation}
Then the equation (\ref{modified gravity}) reads
\begin{equation}
\label{modified gravity1}
\frac{GM}{R^2}=\frac{\pi}{12c\omega_D}a^2=\frac{a^2}{a_0}\ ,
\end{equation}
where the constant $a_0$ is defined as
\begin{equation}
\label{a0}
a_0\equiv\frac{12c\omega_D}{\pi}.
\end{equation}
Thus, $x$ can be rewritten as
\begin{equation}
x= \frac{2\pi c\omega_D}{a}\ = \frac{\pi^2a_0}{6a}\
\end{equation}
and the equation (\ref{Debye function}) turns out to be
\begin{eqnarray}
\mathfrak{D}(x)=\frac{6}{\pi^2}\frac{a}{a_0}\int_0^{\frac{\pi^2a_0}{6a}}\frac{y}{e^y-1}dy\ .
\end{eqnarray}
Defining the function
\begin{equation}
\label{definition of mu}
\mu(t)=\frac{6}{\pi^2}t\int_0^{\frac{\pi^2}{6t}}\frac{y}{e^y-1}dy\ ,
\end{equation}
where $t=a/a_0$, one can immediately obtain the gravity law of MOND\cite{Xin Li}:
\begin{equation}
\label{MOND gravity}
\frac{GM}{R^2}=a\mathfrak{D}\left(\frac{2\pi c\omega_D}{a}\right)=a\mu(x)\ .
\end{equation}
Note that the definition of $\mu(x)$ in the equation (\ref{definition of mu}) has the property\cite{Milgrom}
\begin{equation}
\lim_{x\gg1}\mu(x)=1~~~{\rm and}~~~\lim_{x\ll1}\mu(x)= x\ £¬
\end{equation}
as demanded by MOND.

The equation (\ref{MOND gravity}) implies the modified Poisson equation for the gravitational potential $\phi$
\begin{equation}
\label{Poisson}
\nabla\cdot(\mu(|\nabla\phi|/a_0)\nabla \phi)=4\pi G\rho\ ,
\end{equation}
where $\rho$ is the energy density of matter sources.

In MOND, in order to explain the observed rotational velocity curves of spiral galaxies, $a_0$ is suggested to be of the order of
\begin{equation}
\label{MOND relation}
a_0\sim~10^{-8}cm/s^2\ .
\end{equation}
Milgrom\cite{Milgrom,Milgrom1} found that
\begin{equation}
\label{MOND relation}
2\pi a_0\approx cH_0\ .
\end{equation}
Combining the equation (\ref{a0}) and (\ref{MOND relation}), we get that
\begin{equation}
\label{rel H0 Wd}
H_0\approx24\omega_D\ .
\end{equation}
Thus, the corresponding wave length of $\omega_D$ is
\begin{equation}
\lambda<c/\omega_D=c/24H_0\ .
\end{equation}
It really should be less than the cosmological horizon $L_h\sim c/H_0$. Thus, Milgrom's relation between the threshold of the acceleration $a_0$ and the Hubble constant $H_0$ has a deep origin.

\section{3. The Cosmological Constraints}
\subsection{3.1 The Modified Friedmann Model}
For convenience, we adopt the natural unit $c=k_B=1$ in this section.

The $(3+1)$-demension Friedmann-Robertson-Walker(FRW) metric can be written as
\begin{equation}
 \label{2eq2}
 ds^2 = h_{ab}dx^adx^b +\tilde r^2 d\Omega_{2}^2\ ,
 \end{equation}
 where $\tilde r = a(t) r$ and the
 two-dimension metric $h_{ab} = {\rm diag} (-1, a^2/(1-kr^2))$,  with $k=0,-1,1$
 refers to a flat, open and closed universe respectively.
 Given $h^{ab} \partial_a \tilde r \partial _b \tilde r=0$, one gets the radius of the apparent horizon(AH)\cite{Hawking3}:
\begin{equation}
 \label{2eq3}
 \tilde r_A = \frac{1}{\sqrt{H^2 +k/a^2}}\ .
 \end{equation}
  $H\equiv \dot a/a$ is the Hubble parameter and the dot ``~$\cdot$~'' denotes the derivative with respect to time $t$. The area of the AH is $A= 4\pi {\tilde r_A}^2$. The number of bits of information stored on the AH is
 \begin{equation}
 \label{N}
 N=4\pi\tilde{r}_A^2/L_p^2\ .
\end{equation}

 Suppose from time $t$ to $t+dt$, the radius of the AH changes from $\tilde r_A$ to $\tilde r_A+ d\tilde r_A$. The full derivative of the identity (\ref{modified equipartition}) gives the change of the total energy on the screen as
\begin{equation}
\label{dE}
 dE = \frac{1}{2}N\mathfrak{D}(x)dT+ \frac{1}{2}T \mathfrak{D}(x)dN+ \frac{1}{2}NTd\mathfrak{D}(x)\ .
\end{equation}

Let's first do formal reductions of the above identity.
The temperature of the AH, the so-called Hawking temperature\cite{Hawking2,Cai2}, is
\begin{equation}
\label{T_A}
T_A=\hbar/(2\pi \tilde{r}_A)\ .
\end{equation}
The change of $T_A$ is of the form
\begin{equation}
\label{dT}
dT_A=-\frac{\hbar}{2\pi\tilde{r}_A^2}d\tilde{r}_A\ .
\end{equation}
From the equation (\ref{N}), we get
\begin{equation}
\label{dN}
dN=\frac{8\pi\tilde{r}_A}{L_p^2}d\tilde{r}_A\ .
\end{equation}
The full derivative of the one-dimension Debye function takes the form
\begin{equation}
\label{dD(x)}
d\mathfrak{D}(x)= \left[-\frac{1}{x}\mathfrak{D}(x)+\frac{1}{e^x-1}\right]dx\ .
\end{equation}
The identity (\ref{x}) gives
\begin{equation}
\label{dx}
dx=-\frac{x}{T}dT\ .
\end{equation}
Thus, the equation (\ref{dD(x)}) can be rewritten as
\begin{equation}
\label{dD(x)2}
d\mathfrak{D}(x)= \left[\mathfrak{D}(x)-\frac{x}{e^x-1}\right]\frac{dT_A}{T_A}\ .
\end{equation}

Making use of the equations (\ref{N}), (\ref{T_A}), (\ref{dT}), (\ref{dN}) and (\ref{dD(x)2}
), one can rewrite the equation (\ref{dE}) as
\begin{eqnarray}
\label{dE2}
dE
  =&&\left[\frac{1}{2}\cdot\frac{4\pi\tilde{r}_A^2}{L_p^2}\cdot\left(-\frac{\hbar}{2\pi\tilde{r}_A^2}d\tilde{r}_A\right) +\frac{1}{2}\cdot\left(\frac{\hbar}{2\pi\tilde{r}_A}\right)\cdot\left(\frac{8\pi\tilde{r}_A}{L_p^2}d\tilde{r}_A)\right)\right]\cdot\mathfrak{D}(x)\nonumber\\ &~~&+\frac{1}{2}NT_A\cdot\left[\mathfrak{D}(x)-\frac{x}{e^x-1}\right]\frac{dT_A}{T_A}\nonumber\\
  =&&\left(\frac{x}{e^{x}-1}\right)\frac{d\tilde{r}_A}{G}\ .
\end{eqnarray}
From the definition of the AH (\ref{2eq3}) and remembering $H\equiv\dot{a}/a$, we have
\begin{equation}
\label{draeq}
d\tilde{r}_A=-H\tilde{r}_A^3\left(\dot{H}-\frac{k}{a^2}\right)dt\ .
\end{equation}
Thus, we get
\begin{equation}
\label{draeq}
dE=-H\tilde{r}_A^3\left(\dot{H}-\frac{k}{a^2}\right)\left(\frac{x}{e^{x}-1}\right)\frac{dt}{G}\ .
\end{equation}

Next, we consider the energy-momentum tensor of a perfect isotropic fluid $T_{\mu\nu}=(\rho +p)U_{\mu}U_{\nu} +pg_{\mu\nu}$, where $\rho$ and $p$ are respectively the energy density and the
pressure. This is just the case of the matter in the universe. The energy flow of the matter is\cite{Shu}
\begin{equation}
\label{dE1}
dE =4\pi \tilde{r}_A^2 T_{\mu\nu}k^{\mu}k^{\nu}dt=4\pi\tilde{r}_A^3(\rho+p)Hdt,
\end{equation}
where the Killing vector of the horizon $k^{\mu}=(1,-Hr,0,0)$.

Identifying the equation (\ref{draeq}) and (\ref{dE1}), one obtains
\begin{equation}
\label{FE1}
\left(\frac{x}{e^{x}-1}\right)\left(\dot H -\frac{k}{a^2}\right) =-4\pi G (\rho +p).
\end{equation}

Combining the equation (\ref{FE1}) with the energy conservation equation
\begin{equation}
\dot \rho + 3 H(\rho +p) =0\ ,
\end{equation}
we get
\begin{equation}
\label{FE2}
\left(\frac{x}{e^{x}-1}\right)\left(\dot H -\frac{k}{a^2}\right)=-\frac{4\pi G}{3H}\dot\rho\ .
\end{equation}

Notice that we consider a spatially flat, matter dominated universe throughout this paper, namely $k=0$ and  $\rho=\rho_m=\rho_{m0}\,(1+z)^3=\rho_{m0}\,a^{-3}$. The equation (\ref{FE2}) multiplying by $2Hdt/H_0^2$ makes

\begin{eqnarray}
\label{FE2}
\left(\frac{x}{e^{x}-1}\right)\frac{2H\dot H}{H_0^2} dt &=&-\frac{8\pi G}{3H_0^2}\dot\rho dt\ , \nonumber\\
\left(\frac{x}{e^{x}-1}\right)\frac{d(H^2)}{H_0^2} &=&-\frac{8\pi G}{3H_0^2}d\rho\ .
\end{eqnarray}

With the notations
\begin{equation}
\label{notation}
\Omega_{m0}\equiv\frac{8\pi G\rho_{m0}}{3H_0^2}\,,~~~~~~~
E\equiv\frac{H}{H_0}=\frac{\zeta}{x}\,,~~~~~~~
\zeta\equiv\frac{H_D}{H_0}\equiv \frac{2\pi \omega_D}{H_0}\,,
\end{equation}
one obtains
\begin{equation}
\label{FE3}
\left(\frac{x}{e^x-1}\right)dE^2=\Omega_{m0}~da^{-3}\ .
\end{equation}
$E$ is also called the reduced Hubble parameter.

Instead of numerically solving the differential equation (\ref{FE3}) to get the modified Friedmann equation of the MEF model, we need to find an approximated expression of it. Note that $E(z)$ grows rapidly as the redshift $z$ increases. Suppose that $x=\zeta/E(z)$ is a small quantity today, i.e. $x\ll1$, so that we can expand $x/(e^x-1)$ in powers of $x$ near $x=0$:
\begin{equation}
\label{approximation}
\frac{x}{e^x-1}=1-\frac{x}{2}+ {\cal O}(x^2)~,~~~~~~~~x\ll1\ .
\end{equation}

Then we get the approximated expression of the equation (\ref{FE3})
\begin{eqnarray}
\label{dE2}
\Omega_{m0}~da^{-3}
                   &\thickapprox&\left(2E-\zeta\right)dE\ .
\end{eqnarray}
Integrating the equation (\ref{dE2}), we have
\begin{equation}
\label{FE4}
E^2-\frac{3}{2}\,\zeta E=\Omega_{m0}\,a^{-3}+{\rm const.}\ ,
\end{equation}
where ``~{\rm const.}~'' is an integral constant and should be determined by $E(a=1)=1$. Finally, one obtains a quadratic equation for $E$,
\begin{equation}
\label{FE5}
E^2-\zeta E=\Omega_{m0}a^{-3}+(1-\zeta-\Omega_{m0})\ .
\end{equation}
Note that $E$ only allows of a positive value by its definition (\ref{notation}). Solving the equation (\ref{FE5}), one gets the Friedmann equation in the ODD-MEF model,
\begin{equation}
\label{E2(a)}
E(a)
=\frac{1}{2}\zeta +\frac{1}{2}\left[(\zeta-2)^2+4\Omega_{m0}(a^{-3}-1)\right]^{1/2}\ ,
\end{equation}
or equivalently,
\begin{equation}
\label{E2(z)}
E(z)
=\frac{1}{2}\zeta +\frac{1}{2}\left[(\zeta-2)^2+4\Omega_{m0}((1+z)^{3}-1)\right]^{1/2}\ .
\end{equation}
As long as $\zeta\rightarrow0$, the ODD-MEF model reduces to the $\Lambda$-CDM model, in which
\begin{equation}
E(a)=\left[\Omega_{m0}a^{-3}+\left(1-\Omega_{m0}\right)\right]^{1/2}\ .
\end{equation}

In astronomy, distance modulus is defined as\cite{Carroll}
\begin{equation}
\label{mu}
\mu_{th}(z)\equiv 5\log_{10}[d_L(z)({\rm Mpc})]+25\ ,
\end{equation}
where $d_L$ is the luminosity distance:
\begin{eqnarray}
\label{d_L}
d_L(z)
   &=&\frac{(1+z)}{H_0}\int_{(1+z)^{-1}}^1 \frac{da}{a^2E(a)}\nonumber\\
   &=&\frac{(1+z)}{H_0}\int_{(1+z)^{-1}}^1 \frac{2da}{a^2\cdot\left\{\zeta +\left[(\zeta-2)^2+4\Omega_{m0}(a^{-3}-1)\right]^{1/2}\right\}}\ .
\end{eqnarray}
Finally, we give the specific form of the deceleration parameter $q$ as a function of $E(z)$. Remembering $H\equiv\dot{a}/a$, $a=1/(1+z)$ and
\begin{eqnarray}
\label{d/dt}
\frac{d}{dt}=\frac{da}{dt}\cdot\frac{dz}{da}\cdot\frac{d}{dz}
   =-\frac{H}{a}\frac{d}{dz}\ ,
\end{eqnarray}
one can easily get the following identities
\begin{equation}
\label{q}
q\equiv-\frac{\ddot{a}}{aH^2}
=-1+(1+z)\,E^{-1}\frac{dE}{dz}\ .
\end{equation}

\subsection{3.2 The Numerical Study}
In this subsection, we consider the cosmological constraints on the ODD-MEF model from the observational data. First, we limit ourselves to the 557 SNe Ia from the Union2 compilation of the Supernova Cosmology Project (SCP) collaboration . Next, we include the data from the observation of the large-scale structure (LSS) and the cosmic microwave background (CMB) as useful complements for the SNe Ia data to put a joint constraint on the parameters of the model.

The Union2 compilation consists of 557 SNe Ia, which is the largest published and spectroscopically confirmed sample of SNe Ia so far\cite{SCP}. The $\chi^2$ statistic of the sample is given by
\begin{equation}
\label{chisqSN}
\chi_{SN}^2(\zeta,\Omega_{m0})=\sum\limits_{i=1}^{557}\frac{\left[\mu_{obs}(z_i)-\mu_{th}(z_i;\zeta,\Omega_{m0})\right]^2}{\sigma^2(z_i)}\ ,
\end{equation}
where $\mu_{obs}(z_i)$ and $\sigma(z_i)$ are respectively the observed value and the $1\sigma$ uncertainty of the distance modulus of 557 Union2 SNe Ia. Given the redshift $z_i$ and the parameter values of $\zeta$ and $\Omega_{m0}$, one can get the theoretical, model-related value of the distance modulus $\mu_{th}$ according to the equation (\ref{mu}). By minimizing the $\chi_{SN}^2$, one obtains the best-fit values of the parameters: $\zeta\simeq10^{-7}$ and $\Omega_{m0}=0.192$, with a minimum value of $\chi^2$ as $\chi_{min}^2=574.583$.

\begin{figure}
\begin{center}
\scalebox{1.5}[1.5]{\includegraphics{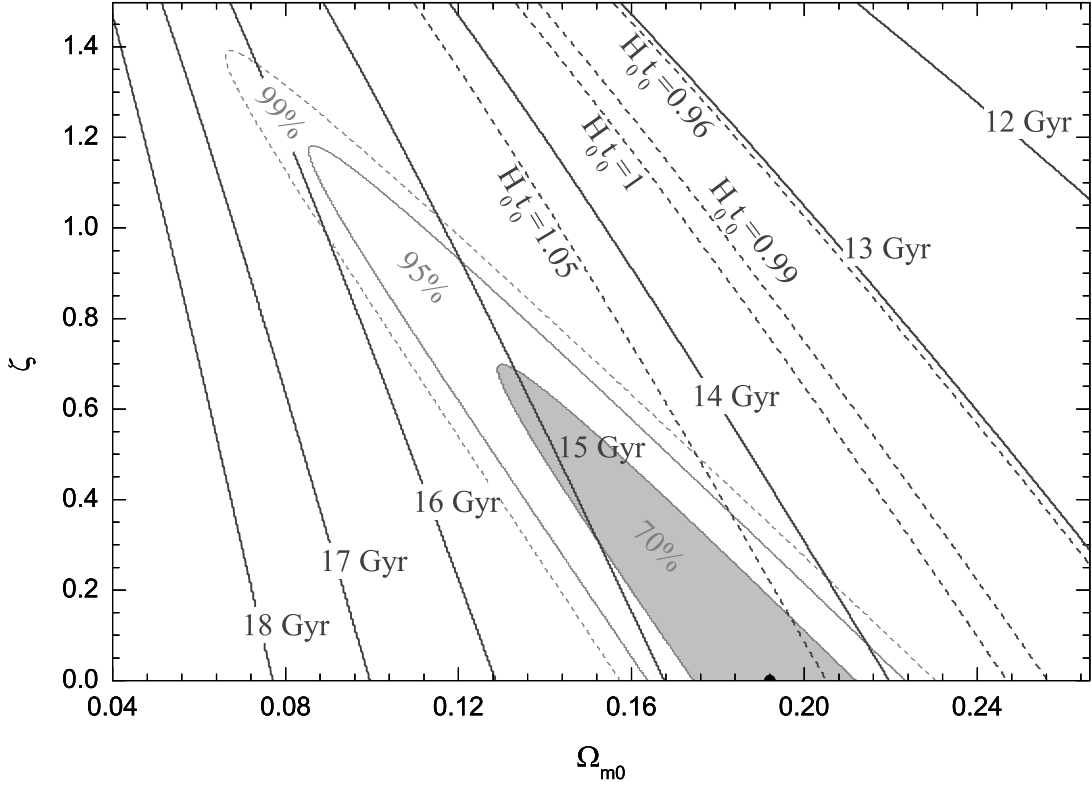}}
\caption{The best-fit $70\%$, $95\%$ and $99\%$ confidence regions in the $\Omega_{m0}$-$\zeta$ plane with the isochrones of constant $H_{0}t_{0}$, for the Union2 compilation of the Supernova Cosmology Project(SCP) collaboration. The dark gray dashed line denotes constraints from the age of the universe at the $1\sigma$ confidence level $0.96\lesssim H_0t_0\lesssim1.05$, with a central value $H_0t_0\simeq0.99$. It can be seen that the best-fit result of the SNe Ia data is almost excluded by the age of the universe at the $1\sigma$ confidence level. The isochrones are labeled for the case of $H_0 = 72~{\rm km~ s^{-1}~Mpc^{-1}}$. The dark gray thick solid lines indicate the age of the universe $t_0$ in the unit of Hubble time $H_{0}^{-1}$. The dot with the coordinate $(0.192, 10^{-7})$ represents the best-fit value, with $\chi_{min}^2=574.583$. The best-fit result is almost excluded by E.~Komatsu {\it et al.}\cite{WMAP7} in $1\sigma$ confidence level.}
\label{MEF-SNIa}
\end{center}
\end{figure}

In addition, we consider the cosmic age in the ODD-MEF model. The age of the universe is given by
\begin{equation}
\label{age}
t_0=H_0^{-1}\int_{0}^{\infty}\frac{dz}{(1+z)E(z)}\ ,
\end{equation}
where $H_0^{-1}$ represents the Hubble time, with the value $H_0^{-1}=9.778h^{-1}$Gyr. We take $h$ to be $0.72$. According to D.~Spergel {\it et al.}\cite{Spergel2003}, the age of the universe is $t_0=13.7\pm0.2$Gyr. This can also be written in the dimensionless age parameter $H_0t_0\simeq0.99$, with $1\sigma$ confidence level range $0.96\lesssim H_0t_0\lesssim1.05$\cite{Zhang}. We show these in FIG.\ref{MEF-SNIa}, together with the $70\%$, $95\%$ and $99\%$ confidence level contours in the $\Omega_{m0}$-$\zeta$ parameter plane.

As we see, the SNe Ia data alone are not enough to put a very strong constraint on the parameters. The parameter $\zeta$ can still take a value close to $1$. This fact invalidates the approximation (\ref{approximation}). To recognize this, one should notice that $E(z)\equiv H/H_0=\zeta/x=1$ today. A large $x$ implies a large $\zeta$. Moreover, one can see that the best-fit result of the SNe Ia data is almost ruled out by the age of the universe at $1\sigma$ confidence level ($0.96\lesssim H_0t_0\lesssim1.05$). So in what follows, we use the data from the observation of the large-scale structure (LSS) and the cosmic microwave background (CMB) to enforce a more rigorous constraint on the model parameters.

Here, we only take into account the most conservative and robust informations from the LSS and the CMB observations\cite{Zhang}. For the LSS, we use the distance parameter ${\cal A}$. It can be obtained from a spectroscopic sample of 46,748 luminous red galaxies of the Sloan Digital Sky Survey (SDSS). It is defined as\cite{SDSS,Eisenstein}
\begin{equation}
{\cal A}\equiv\Omega_{m0}^{1/2}\,E(z_b)^{-1/3}\left[\frac{1}{z_b} \int_0^{z_b}\frac{dz}{E(z)}\right]^{2/3}\ ,
\label{A}
\end{equation}
where $z_b=0.35$. $E(z)$ is given by equation (\ref{E2(z)}). The value of ${\cal A}$ has been determined to be ${\cal A}_0=0.469\pm\sigma_{\cal A}$, where the $1\sigma$ uncertainty $\sigma_{\cal A}=0.017$\cite{Eisenstein}.

For the CMB, we use the measurement of shift parameter $\cal R$ alone. It is defined as\cite{Wang,Tegmark97}
\begin{equation}
\label{R}
{\cal R}\equiv\Omega_{m0}^{1/2}\int_0^{z_\ast}{dz\over E(z)}~,
\end{equation}
where the redshift of recombination is found to be ${z_\ast}=1091.3$ according to the WMAP 7-year (WMAP7) data\cite{WMAP7}. The value of $\cal R$ has been updated to ${\cal R}_0=1.725\pm\sigma_{\cal R}$, where the $1\sigma$ uncertainty $\sigma_{\cal R}=0.018$\cite{WMAP7}.

\begin{figure}
\begin{center}
\scalebox{1.5}[1.5]{\includegraphics{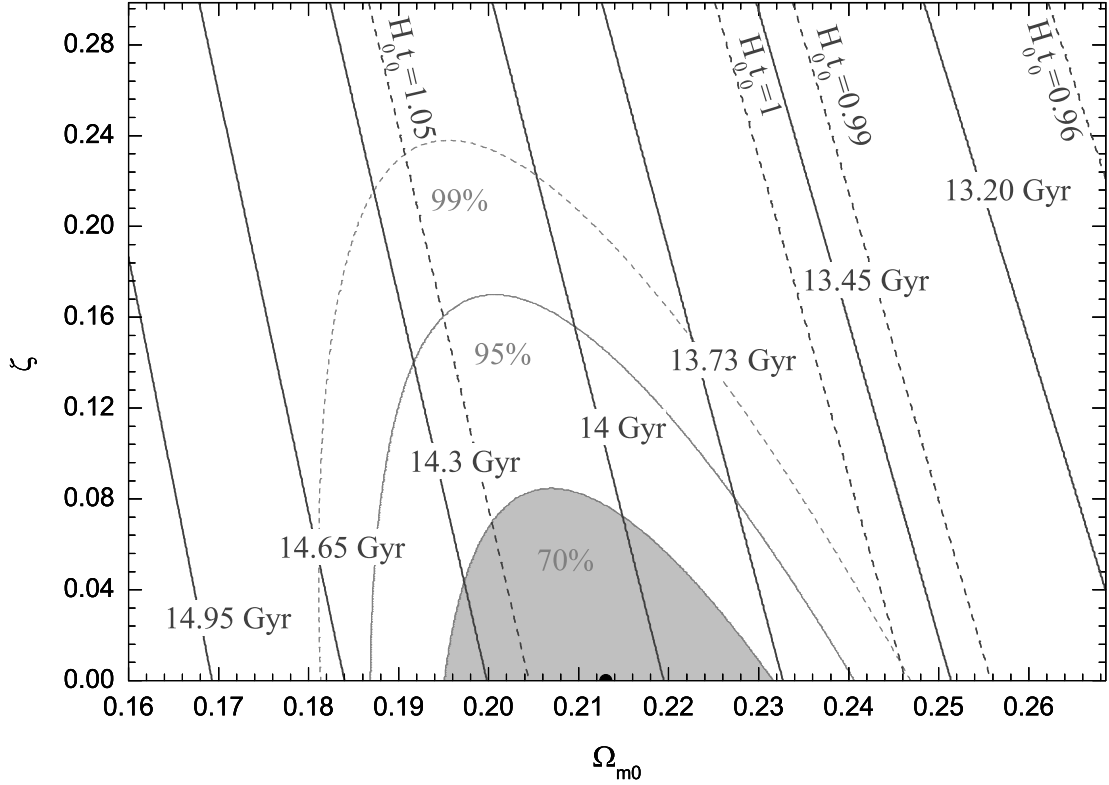}}
\caption{The best-fit $70\%$, $95\%$ and $99\%$ confidence regions in the $\Omega_{m0}$-$\zeta$ plane with the isochrones of constant $H_{0}t_{0}$, for the Union2 compilation of the Supernova Cosmology Project(SCP) collaboration and the data from the observation of the large-scale structure(LSS), i.e. SNe Ia+LSS. The dark gray dashed lines denote constraints from the age of the universe at $1\sigma$ confidence level $0.96\lesssim H_0t_0\lesssim1.05$, with a central value $H_0t_0\simeq0.99$. The isochrones are labeled for the case of $H_0 = 72~{\rm km~ s^{-1}~Mpc^{-1}}$.  The dark gray thick solid lines indicate the age of the universe $t_0$ in the unit of Hubble time $H_{0}^{-1}$. The dot with the coordinate $(0.213, 10^{-8})$ represents the best-fit value, with $\chi_{min}^2=584.831$. The $95\%$ confidence level of the parameter $\zeta$, i.e. $0\leq \zeta \leq 0.17$, is much smaller than the SNe Ia-only case.}
\label{MEF-SNIa+LSS}
\end{center}
\end{figure}

To include both the LSS and the CMB observation as well as the 557 Union2 SNe Ia data to perform a combined numerical analysis of the parameters, we use a $\chi^2$ statistic as
\begin{equation}
\chi^2=\chi_{SN}^2+\chi_{LSS}^2+\chi_{CMB}^2\ ,
\end{equation}
where $\chi_{SN}^2$ is given by the identity (\ref{chisqSN}). $\chi_{CMB}^2$ and
$\chi_{LSS}^2$ are the contributions from the CMB and the LSS data, which are defined as
\begin{equation}
\label{chisqLSS}
\chi^2_{LSS}=\frac{({\cal A}-{\cal A}_0)^2}{\sigma_{\cal A}^2}~~~and~~~\chi^2_{CMB}=\frac{({\cal R}-{\cal R}_0)^2}{\sigma_{\cal R}^2}\ ,
\end{equation}
where ${\cal A}$ and $\cal R$ are given by the identity (\ref{A}) and (\ref{R}) respectively. By minimizing such a $\chi^2$, one gets the best-fit parameter values $\zeta\simeq10^{-9}$ and $\Omega_{m0}=0.224$, with $\chi_{min}^2=591.156$. These results are shown in FIG.\ref{MEF-SNIa+LSS+CMB}. For comparison, we also present the result for the SNe Ia+LSS fit in FIG.\ref{MEF-SNIa+LSS}, of which we use a $\chi^2$ statistic
\begin{equation}
\chi^2=\chi_{SN}^2+\chi_{LSS}^2\ .
\end{equation}
The best-fit parameter values are $\zeta\simeq10^{-8}$ and $\Omega_{m0}=0.213$, with $\chi_{min}^2=584.831$.

One can see that after using the LSS and the CMB data, we successfully put a much more rigorous constraint on $\zeta$. It is within the range $0\leq \zeta \leq 0.09$ at $95\%$ confidence level. This result confirms that our approximation (\ref{approximation}) is reasonable. And it is more than one order smaller than Wei's prediction\cite{Wei}. In addition, the best-fit result of the combined analysis is in accordance with the age of the universe given by D.~Spergel {\it et al.}\cite{Spergel2003} at $1\sigma$ confidence level.

\begin{figure}
\begin{center}
\scalebox{1.65}[1.65]{\includegraphics{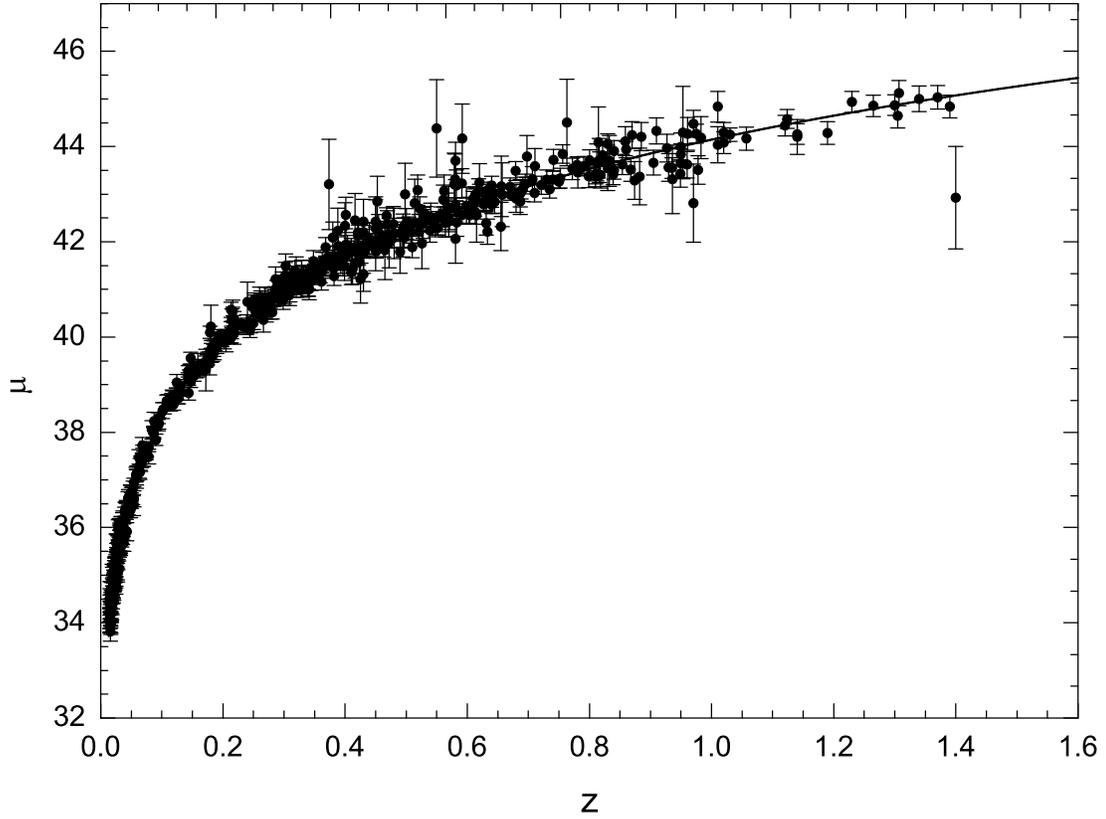}}
\caption{The distance modulus $\mu$ versus redshift $z$ for the Union2 compilation of the Supernova Cosmology Project(SCP) collaboration plus the data from the observation of the large-scale structure(LSS) and the cosmic microwave background(CMB), i.e. SNe Ia+LSS+CMB. The observation consists of 557 SNIa data points with $1\sigma$ error bars. The theoretically computed result $\mu_{th}$ with the best-fit parameter values $\zeta=10^{-9}$ and $\Omega_{m0}=0.224$ are plotted in black solid line for the one-dimension Debye model of modified entropic force.}
\label{mu(z)}
\end{center}
\end{figure}

\begin{figure}
\begin{center}
\scalebox{1.6}[1.6]{\includegraphics{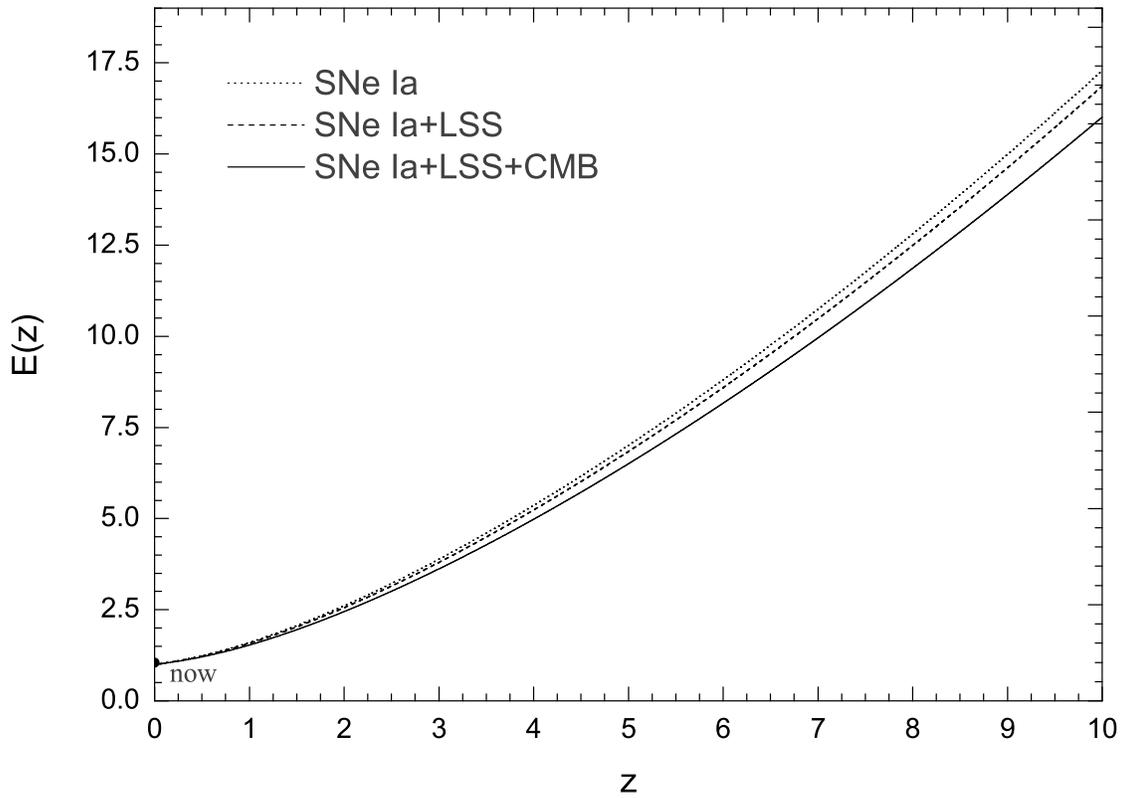}}
\caption{The reduced Hubble parameter $E$ versus redshift $z$ for three fits, namely SNe Ia-only, SNe Ia+LSS, and SNe Ia+LSS+CMB, denoted by the dark gray dotted line, dashed line, and the solid line, respectively. It can be seen that $E(z)$ grows rapidly as the redshift $z$ increases.}
\label{E(z)}
\end{center}
\end{figure}

\begin{figure}
\begin{center}
\scalebox{1.6}[1.6]{\includegraphics{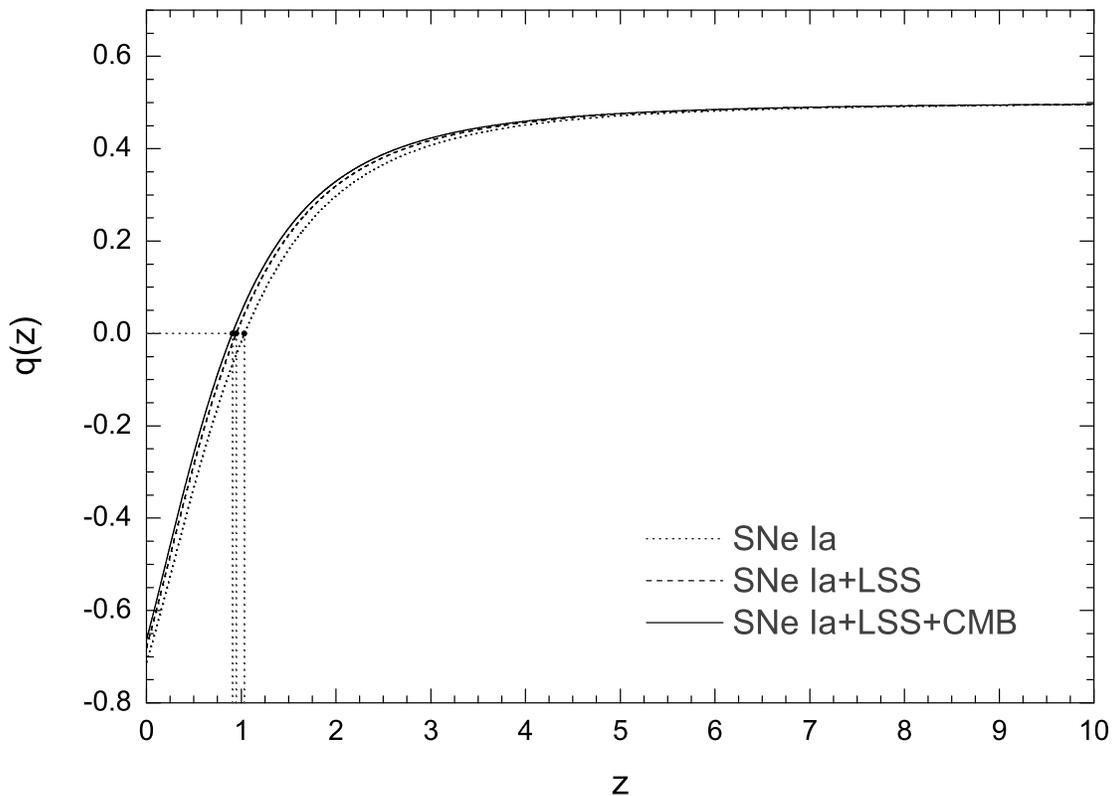}}
\caption{The deceleration parameter $q$ versus redshift $z$ for three fits, namely SNe Ia-only, SNe Ia+LSS, and SNe Ia+LSS+CMB. Note that $q=0$ at the redshift $z_t=1.034$ for SNe Ia only, $z_t=0.948$ for SNe Ia+LSS, and $z_t=0.906$ for SNe Ia+LSS+CMB. $z_t$ is called the transition redshift, which implies that the universe goes from a decelerated expanding period to an accelerated expanding period at a time point in the early epoch.}
\label{q(z)}
\end{center}
\end{figure}

The corresponding distance modulus $\mu_{th}(z)$ is shown in FIG.\ref{mu(z)}. And the corresponding reduced Hubble parameter $E(z)$ and the deceleration parameter $q(z)$ are plotted in FIG.\ref{E(z)} and FIG.\ref{q(z)}. Notice that $q=0$ at the redshift $z_t=0.906$ for the SNe Ia+LSS+CMB fit. $z_t$ is called the transition redshift. This fact implies that the universe goes from a decelerated expanding period to an accelerated expanding period at a time point in the early epoch, in the absence of any dark matter or dark energy.



\section{4. The Pioneer 10/11 Anomaly}
Last, but not the least, let us consider the Pioneer 10/11 anomaly in the ODD-MEF model. The Pioneer 10 and 11 spacecraft, which were launched in 1972 and 1973, are the two most accurately navigated vehicles in the solar system. The radio-metric data received by them at the heliocentric distances ranging from $20$-$70$ AU has consistently indicated the existence of a small, anomalous frequency drift changing with a rate of $6\times10^{-9}$ Hz/s\cite{Anderson}, which was later interpreted as a constant, sunward acceleration\cite{Turyshev1}
\begin{equation}
\label{a_P}
a_P=(8.74 \pm1.33)\times10^{-10}~m/s^2\ .
\end{equation}
This apparent violation of the Newton's law of gravity has been known as the Pioneer anomaly, the nature of which still remains unclear to date.

Several mechanisms have been developed to explain for the anomaly. To name a few\cite{Turyshev2}, it could be attributed to the possible systematic errors due to the gas leaks from the propulsion system, the gravitational effect of the Kuiper Belt Objects or dust, or the influence of the expansion of the universe, etc. In our previous work\cite{Xin Li2}, we have also provided a possible explanation of the Pioneer 10/11 anomaly in the framework of Finsler geometry.

Here, we present another possible interpretation of the Pioneer anomaly in the ODD-MEF model. Starting from the identity (\ref{modified gravity}) and using the approximation (\ref{approximation}), we have
\begin{eqnarray}
\label{a}
\frac{GM}{R^2}=a\mathfrak{D}(x)&=&a\cdot\frac{1}{x}\int^x_0\frac{y}{e^y-1}dy\nonumber\\
&=&a-\frac{\pi cw_D}{2}\ .
\end{eqnarray}
We may rewrite it as
\begin{equation}
\label{a2}
a=\frac{GM}{R^2}+\frac{\pi cw_D}{2}\ .
\end{equation}

In the above identity, $GM/R^2$ is the conventional Newtonian prediction of the gravitational acceleration at a distance $R$ from the source with mass $M$. We take the left hand side of the equation (\ref{a2}), i.e. $a$,  to be the acceleration that the spacecraft actually experiences. Then the second term at the right hand side should be interpreted as the abnormal acceleration $a_P$, namely
\begin{equation}
\label{a3}
a=\frac{GM}{R^2}+a_P\ ,
\end{equation}
where
\begin{equation}
\label{a_P2}
a_P\equiv\frac{\pi cw_D}{2}\ .
\end{equation}

Suppose that the parameter takes a value as $\zeta\approx5$ over the heliocentric distance ranging from $20$ to $70$ AU. This is the structure scale where the abnormal acceleration $a_P$ has been detected. Using the definition (\ref{notation}) of $\zeta$, we get the magnitude of $\omega_D$
\begin{equation}
\label{w_D}
w_D=\frac{H_0}{2\pi}\zeta\approx2.0\times10^{-18}~{\rm Hz}\ .
\end{equation}
With the identity (\ref{a_P2}), one gets the abnormal acceleration predicted by the ODD-MEF model
\begin{equation}
\label{a_P3}
a_P\equiv\frac{\pi cw_D}{2} ~\approx ~8.74\times10^{-10}~m/s^2\ .
\end{equation}

It agrees with the detected result (\ref{a_P}). The plus sign in the equation (\ref{a3}) illustrates that the direction of $a_P$ is in the same direction as the conventional Newtonian prediction $GM/R^2$, which is pointing towards the sun. This is in accordance with the detection.


\section{5. A Unified Scenario of Anomalies in Gravity}
In section 4, we found that $\zeta\approx5$ provides a possible explanation for the Pioneer 10/11 anomaly in the solar system. On the other hand, to account for the observations of the SNe Ia, the LSS and the CMB, $\zeta$ is found to be $\zeta\sim10^{-9}$. This was demonstrated in section 3.

In the following, we will find out the magnitude of $\zeta$ corresponding to MOND.
Making use of
\begin{equation}
\label{w_D2}
w_D=\frac{H_0}{2\pi}\zeta
\end{equation}
and the identity (\ref{a0}), we get
\begin{eqnarray}
\label{a0_2}
a_0
   &=&\frac{6cH_0}{\pi^2}\zeta\ .
\end{eqnarray}
The above identity multiplying $2\pi$ makes
\begin{equation}
\label{2pia}
2\pi a_0=\frac{12\zeta}{\pi}cH_0\ .
\end{equation}
Note that in the equation (\ref{MOND relation}),~$2\pi a_0\approx cH_0$. So we have
\begin{equation}
\frac{12\zeta}{\pi}\approx 1\ ,
\end{equation}
which implies
\begin{equation}
\zeta\approx 0.25\ .
\end{equation}

In summary, we obtain three different magnitudes of $\zeta$ with respect to three different structure scales:
\begin{equation}\zeta\sim
\label{zeta}
\begin{cases}10^{-9}\ , & 10^{8}\sim10^{10}~$pc~~~~~$(\text{SNe Ia+LSS+CMB}) \\ 0.25\ , & 10^{3}\sim10^{4}~$pc~~~~~~$(\text{MOND}) \\ 5\ , & 10^{-4}\sim10^{-3}~$pc~~~$(\text{Pioneer 10/11 anomaly})
\end{cases}
\ ~~~,
\end{equation}
where $1~$pc$=2.06\times10^{5}$~AU$=3.09\times10^{16}$~m$=3.26$~l.y.. One can see that $\zeta$ is found to be of a declining value while the relevant structure scale increases.

\section{6. Conclusions and Remarks}
Verlinde\cite{Verlinde} interpreted gravity as an entropic force due to the tendency of the system to increase its entropy, which is associated with the change of positions of material bodies. With the holographic principle, the total entropy of the system can be calculated via the area $A$ and the temperature $T$ of its boundary. With Unruh's relation of the acceleration and the temperature experienced by the accelerated observer, one can easily recover the laws of Newton. However, at very low temperatures, the equipartition law of energy is no longer valid. We noted that the threshold of low temperature is related to the specific dimension of the direction. In the case of astronomy and cosmology, the only threshold comes from the limited radius of the solar system , the galaxies and the universe. We made a modification by revising the equipartition law of energy with the one-dimension Debye function (\ref{Debye function}). It was referred as the ODD-MEF model.

\begin{figure}
\begin{center}
\scalebox{1.5}[1.5]{\includegraphics{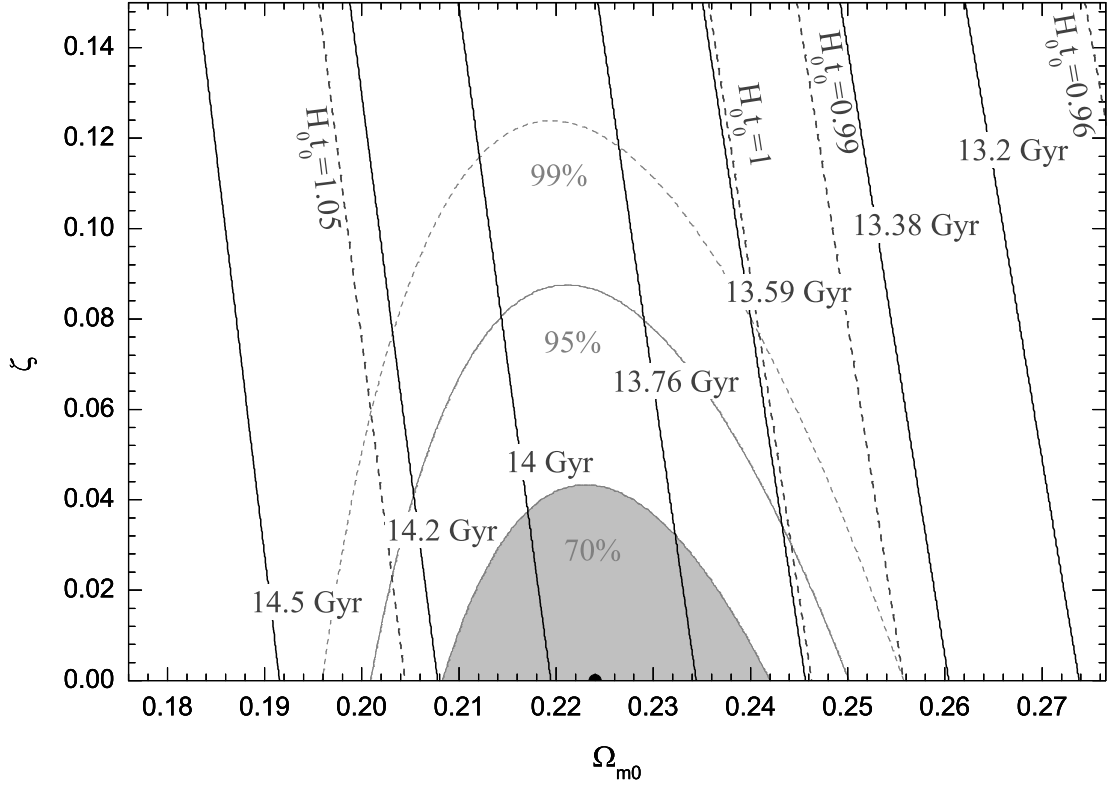}}
\caption{The best-fit $70\%$, $95\%$ and $99\%$ confidence regions in the $\Omega_{m0}$-$\zeta$ plane with the isochrones of constant $H_{0}t_{0}$, for the Union2 compilation of the Supernova Cosmology Project(SCP) collaboration plus the data from the observation of the large-scale structure(LSS) and the cosmic microwave background(CMB), i.e. SNe Ia+LSS+CMB. The dark gray dashed lines denote constraints from the age of the universe at $1\sigma$ confidence level $0.96\lesssim H_0t_0\lesssim1.05$, with a central value $H_0t_0\simeq0.99$. The isochrones are labeled for the case of $H_0 = 72~{\rm km~ s^{-1}~Mpc^{-1}}$.  The dark gray thick solid lines indicate the age of the universe $t_0$ in the unit of Hubble time $H_{0}^{-1}$. The dot with the coordinate $(0.224, 10^{-9})$ represents the best-fit value, with $\chi_{min}^2=591.156$. The best-fit result is in accordance with D.~Spergel{\it et al.}\cite{Spergel2003} at $1\sigma$ confidence level. A small $\zeta$, $0\leq \zeta \leq 0.09$ within $95\%$ confidence level, is obtained.}
\label{MEF-SNIa+LSS+CMB}
\end{center}
\end{figure}

With the definition (\ref{a0}), we deduced the famous formula of MOND. To consider the cosmological constraints on the model, we derived the Friedmann equation (\ref{E2(z)}). We made use of the Union2 compilation of the Supernova Cosmology Project(SCP) collaboration, the data from the observation of the large-scale structure(LSS) and the cosmic microwave background(CMB), i.e. SNe Ia+LSS+CMB, to put a joint constraint on the parameters $\zeta$ and $\Omega_{m0}$. The combined numerical analysis gives the best-fit values of the model parameters $\zeta\simeq10^{-9}$ and $\Omega_{m0}=0.224$, with $\chi_{min}^2=591.156$. In FIG.\ref{MEF-SNIa+LSS+CMB}, one can see that the best-fit result agrees with the age prediction of the universe by D.~Spergel {\it et al.}\cite{Spergel2003} and the WMAP collaboration at $1\sigma$ confidence level. And the parameter $\zeta$ is successfully limited to a narrow range $0\leq \zeta \leq 0.09$ at $95\%$ confidence level. It is more than one order smaller than Wei's prediction $0\leq \zeta \leq 0.2$\cite{Wei}.

In the ODD-MEF model, we introduced a dimensionless, structure scale-related parameter $\zeta$. Phenomenological study indicated that different values of $\zeta$ corresponds to dynamics at different distance scales, as in the identity (\ref{zeta}). Therefore, we are allowed to suggest a unified scenario of the asymptotic flatness of the rotation curves of spiral galaxies, an accelerated expanding universe, and the Pioneer 10/11 anomaly in the solar system in a single framework. Thus, we got a possible unification of dark matter and dark energy in the one-dimensional Debye model of the modified entropic force.

Since $\zeta$ is a running parameter associated with the structure scale $r$, a specific form of the function $\zeta(r)$ should be given. As one foresees, this function should have the monotonically decreasing property as shown in (\ref{zeta}).  Such a running $\zeta$ also implies that, other than the ODD-MEF model, there exists a more fundamental theory, which taking the model as its effective scenario. To find the specific $\zeta(r)$ and such a fundamental theory is one of the main subjects of our future investigation.

\begin{acknowledgments}
\section{Acknowledgments}
This work was supported by the National Natural Science Fund of China under Grant No. 10575106 and No. 10875129.
\end{acknowledgments}

\bibliography{basename of .bib file}

\begin{thebibliography}{999}
\bibitem{Hawking}J. M. Bardeen, B. Carter and S. W. Hawking, Commun. Math. Phys. {\bf 31}, 161 (1973).
\bibitem{Bekenstein}J. D. Bekenstein, Phys. Rev. D {\bf 7}, 949 (1973). J.D. Bekenstein, Phys. Rev. D {\bf 7}, 2333 (1973).
\bibitem{Hawking2}S. W. Hawking, Commun. Math. Phys. {\bf 43}, 199 (1975) [Erratum-ibid. {\bf 46}, 206 (1976)].
\bibitem{Cai2}R. G. Cai, L. M. Cao, and Y. P. Hu, Class.Quant.Grav. {\bf 26}, 155018 (2009).
    arXiv:hep-th/0809.1554.
\bibitem{Jacobson}T. Jacobson, Phys. Rev. Lett. {\bf 75}, 1260 (1995).
\bibitem{Padmanabhan1}T. Padmanabhan, Class. Quantum Grav. {\bf 21}, 4485 (2004).
\bibitem{Padmanabhan2}T. Padmanabhan, ``Thermodynamical Aspects of Gravity:~New Insights," arXiv:gr-qc/0911.5004.
\bibitem{Verlinde}E. P. Verlinde, ``On the Origin of Gravity and the Lawsof Newton," arXiv:hep-th/1001.0785.
\bibitem{Unruh}W. G. Unruh,  Phys. Rev. D {\bf 14}, 870 (1976).
\bibitem{Komar}A. Komar, Phys. Rev. {\bf 113}, 934 (1959).
\bibitem{Susskind}L. Susskind, J. Math. Phys. {\bf 36}, 6377 (1995). G. t Hooft, arXiv:gr-qc/9310026. E. Witten, Adv.
Theor. Math. Phys. {\bf 2}, 253 (1998).
\bibitem{Shu}F. W. Shu and Y. Gong, ``Equipartition of energy and the first law of thermodynamics at the apparent horizon, " arXiv:gr-qc/1001.3237.
\bibitem{Cai1}R. G. Cai, L. M. Cao and N. Ohta, ``Friedmann Equations from Entropic Force," arXiv:hep-th/1001.3470.
\bibitem{Cai3}R. G. Cai, S. P. Kim, and Y. P. Hu, JHEP. {\bf 2}, 50 (2005). arXiv:hep-th/0809.1554.
\bibitem{Cai4}R. G. Cai, L. M. Cao, and Y. P. Hu, JHEP. {\bf 8}, 90 (2008). arXiv:hep-th/0807.1232.
\bibitem{Smolin}L. Smolin, ``Newtonian gravity in loop quantum gravity," arXiv:gr-qc/1001.3668.
\bibitem{Makela}J, M\"{a}kel\"{a},``Notes Concerning `On the Origin of
Gravity and the Laws of Newton' by E. Verlinde," arXiv:gr-qc/1001.3808.
\bibitem{Caravelli}F. Caravelli and L. Modesto, ``Holographic actions from
black hole entropy," arXiv:gr-qc/1001.4364.
\bibitem{LiM}M. Li and Y. Wang, ``Quantum UV/IR Relations
and Holographic Dark Energy from Entropic Force," arXiv:hep-th/1001.4466.
\bibitem{Tian}Y. Tian and X. N. Wu, ``Thermodynamics of Black Holes from Equipartition of Energy
and Holography," arXiv:hep-th/1002.1275.
\bibitem{Gao}C. J. Gao, Phys. Rev. D {\bf 81}, 087306 (2010).
\bibitem{Xin Li}X. Li and Z. Chang, ``Debye entropic force and modified Newtonian dynamics," arXiv:hep-th/1005.1169v2.
\bibitem{Trimble}V. T. Trimble, Ann. Rev. Astron. Astrophys. {\bf 25}, 425 (1987).
\bibitem{Rubin}V. C. Rubin, W. K. Ford, and N. Thonnard, Astrophys. J. {\bf 238}, 471 (1980).
\bibitem{TF}R. B. Tully and J. R. Fisher, Astr. Ap. {\bf 54}, 661 (1977).
\bibitem{Debye}P. Debye, Ann. Physik, {\bf 39}, 789 (1912).
\bibitem{Milgrom}M. Milgrom, Astrophys. J. {\bf 270}, 365 (1983).
\bibitem{Milgrom1}M. Milgrom, ``The MOND paradigm," arXiv:astro-ph/0801.3133.
\bibitem{Hawking3}S. W. Hawking and G.F.R.Ellis, {\it The large scale structure of space-time}. Cambridge
University Press: London, 1973.
\bibitem{Carroll}S. Carroll, ``The Cosmological Constant," Living Rev. Relativity {\bf 4}, 1 (2001).
\bibitem{SCP}R.~Amanullah {\it et al.} [Supernova Cosmology Project Collaboration], arXiv:1004.1711 [astro-ph.CO].\\
 The numerical data of the full Union2 sample are available at http:$/\!/$supernova.lbl.gov/Union
\bibitem{Spergel2003}D. N. Spergel, {\it et al}., Astrophys.J.Suppl. {\bf 148}, 175 (2003).
\bibitem{Zhang}X. Zhang and F. Q. Wu, Phys. Rev. D {\bf 72}, 043524 (2005).
\bibitem{SDSS}
M.~Tegmark {\it et al.} [SDSS Collaboration],
 Phys.\ Rev.\ D {\bf 69}, 103501 (2004) arXiv:astro-ph/0310723;\\
M.~Tegmark {\it et al.} [SDSS Collaboration],
 Astrophys.\ J.\  {\bf 606}, 702 (2004) arXiv:astro-ph/0310725;\\
U.~Seljak {\it et al.} [SDSS Collaboration],
 Phys.\ Rev.\ D {\bf 71}, 103515 (2005) arXiv:astro-ph/0407372;\\
M.~Tegmark {\it et al.} [SDSS Collaboration],
 Phys.\ Rev.\  D {\bf 74}, 123507 (2006) arXiv:astro-ph/0608632.
\bibitem{Eisenstein}D.~J.~Eisenstein {\it et al.} [SDSS Collaboration], Astrophys.\ J.\ {\bf 633}, 560 (2005) arXiv:astro-ph/0501171.

\bibitem{Wang}Y.~Wang and P.~Mukherjee, Astrophys.\ J.\ {\bf 650}, 1 (2006) arXiv:astro-ph/0604051.
\bibitem{Tegmark97}J.~R.~Bond, G.~Efstathiou and M.~Tegmark, Mon.\ Not.\ Roy.\ Astron.\ Soc.\ {\bf 291}, L33 (1997)
 [astro-ph/9702100].
\bibitem{WMAP7}E.~Komatsu {\it et al.} [WMAP Collaboration], arXiv:1001.4538 [astro-ph.CO].
\bibitem{Wei}H. Wei, ``Cosmological Constraints on the Modified Entropic Force Model,'' arXiv:gr-qc/1005.1445v1.

\bibitem{Anderson}J. D. Anderson, {\it et al}., Phys. Rev. Lett. {\bf 81} 2858 (1998), J. D. Anderson, {\it et al}.,
Phys. Rev. D {\bf 65} 082004 (2002), J. D. Anderson, {\it et al}., Mod. Phys. Lett. A {\bf 17} 875 (2002).
\bibitem{Turyshev1}S. G. Turyshev, {\it et al}., Stanford e-Conf \#C041213, \#0310, arXiv:gr-qc/0503021.
\bibitem{Turyshev2}S. G. Turyshev, {\it et al}., EAS Publ. Ser. {\bf 20}, 243 (2006).
\bibitem{Xin Li2}X. Li and Z. Chang, ``A possible scenario of the Pioneer anomaly in the framework of Finsler geometry," arXiv:gr-qc/0909.3713v1.






\end{thebibliography}

\newpage

\newpage

\newpage

\newpage

\newpage

\newpage

\end{document}